\documentclass[12pt]{article}

\usepackage{epsfig}
\input epsf

\title{Prompt photon hadroproduction at high energies\\ in the $k_T$-factorization approach}

\author{A.V.~Lipatov, N.P.~Zotov}

\begin{document}

\maketitle

\begin{center}

{\it D.V.~Skobeltsyn Institute of Nuclear Physics,\\ 
M.V. Lomonosov Moscow State University,
\\119992 Moscow, Russia\/}\\[3mm]

\end{center}

\vspace{1cm}

\begin{center}

{\bf Abstract }

\end{center}

We consider the prompt photon production at high energy hadron colliders
in the framework of $k_T$-factorization approach. 
The unintegrated quark and gluon distributions in a proton are determined 
using the Kimber-Martin-Ryskin prescription. 
The conservative error analisys is performed.
We investigate both inclusive prompt photon and prompt photon and 
associated muon production rates. In Standard Model such events come 
mainly due to Compton scattering process where the final 
heavy (charm or bottom) quark produces a muon.
The theoretical results are compared with recent
experimental data taken by the D$\oslash$ and CDF collaborations at 
Fermilab Tevatron. Our analysis also covers the azimuthal correlations
between produced prompt photon and muon which can provide an important 
information about non-collinear parton evolution in a proton.
Finally, we extrapolate the theoretical predictions to CERN LHC energies.

\vspace{1cm}

\section{Introduction} \indent 

It is well known that production of prompt (or direct) photons at
high energies has provided a direct probe of the hard subprocess dynamics, 
since produced photons are largely insensitive to the final-state 
hadronization effects. Hadroproduction of prompt photons has been studied in a number
of experiments~[1--5]. Usually photons are called "prompt" if they 
are coupled to the interacting quarks. In the framework of Quantum
Chromodynamics (QCD) the dominant production mechanism for the prompt photons
at Tevatron and LHC colliders is the Compton 
scattering $g q \to \gamma q$~[6]. It is clear that cross section of 
such processes is sensitive to the gluon distributions in a proton.
Also observed final state photon may arise from so called fragmentation
processes~[7], where a quark or gluon is transformed into $\gamma$. 
The cross sections of such processes involve relative poorly known
parton-to-photon fragmentation functions~[8].
However, the isolation criterion which is usually introduced in experimental analyses 
substantially reduces the fragmentation component (see, for example, Ref.~[9]).
Therefore for the theoretical description of prompt photon production
at Tevatron the detailed knowledge of parton (quark and gluon) 
distributions in a proton is necessary.

Usually quark and gluon densities in a proton are described by the 
Dokshitzer-Gribov-Lipatov-Altarelli-Parizi (DGLAP) evolution equation~[10]
where large logarithmic terms proportional to $\ln \mu^2$ are taken into 
account only. The cross sections can be rewritten in terms of process-dependent
hard matrix elements convoluted with universal quark and/or gluon density functions.
In this way the dominant contributions come from diagrams where parton emissions 
in initial state are strongly ordered in virtuality. This is called collinear 
factorization, as the strong ordering means that the virtuality of the parton 
entering the hard scattering matrix elements can be neglected compared to the 
large scale $\mu$.

However, at high energies (or small $x \sim \mu^2/s \ll 1$) effects of 
finite virtualities and transverse 
momenta of the incoming partons may become more and
more important. These effects can be systematically accounted for in the 
$k_T$-factorization QCD approach~[10--14]. Just as for DGLAP, in this way it is possible to 
factorize an observable into a convolution of process-dependent 
hard matrix elements with universal parton distributions. 
But as the virtualities (and transverse momenta) of the emitted partons are no 
longer ordered, the matrix elements have to be taken off-shell and the 
convolution made also over transverse momentum ${\mathbf k}_T$ with the 
unintegrated (i.e.~$k_T$-dependent) parton distributions.
The unintegrated parton distribution $f_a(x,{\mathbf k}_T^2)$ 
determines the probability to find a type $a$ parton carrying the 
longitudinal momentum fraction $x$ and the transverse momentum ${\mathbf k}_T$.
In particular, usage of the unintegrated parton densities have the advantage 
that it takes into account true kinematics of the process under consideration 
even at leading order.

The unintegrated parton distributions $f_a(x,{\mathbf k}_T^2)$ are a subject 
of intensive studies~[15, 16]. Various approaches to investigate these quantities 
has been proposed. It is believed that at assymptotically large energies 
(or very small $x$) the theoretically correct description is given by 
the Balitsky-Fadin-Kuraev-Lipatov (BFKL) evolution equation~[17] where 
large terms proportional to $\ln 1/x$ are taken into account. Another 
approach, valid for both small and large $x$, have been developed 
by Ciafaloni, Catani, Fiorani and Marchesini, and is known as the CCFM model~[18].
It introduces angular ordering of emissions
to correctly treat gluon coherence effects. In the limit of 
asymptotic energies, it almost equivalent to the BFKL~[19--21], but also similar to
the DGLAP evolution for large $x \sim 1$. The resulting unintegrated
gluon distribution depends on two scales, the additional scale ${\bar q}$
is a variable related to the maximum angle allowed in the emission and plays 
the role of the evolution scale $\mu$ in the collinear parton densities.

The two-scale involved unintegrated parton 
distributions it is possible to obtain also from the conventional 
ones using the Kimber-Martin-Ryskin (KMR) 
prescription~[22]. In this way the $\mu$ dependence in the unintegrated parton 
distribution enters only in last step of the evolution, and single scale 
evolution equations can be used up to this step. Such procedure can be 
applied to a proton as well as photon and is expected to account for the 
main part of the conventional next-to-leading logarithmic QCD corrections. 
The KMR-constructed parton densities were used, in particular, to describe the 
heavy quark production in $\gamma \gamma$ collisions at CERN LEP2~[23], prompt 
photon photoproduction at DESY HERA~[24] (both inclusive and associated with hadronic 
jet) and inclusive prompt photon hadroproduction at Fermilab Tevatron~[25].

Recently new experimental data on the $p + \bar p \to \gamma + X$ process
at Tevatron were obtained by the D$\oslash$~[1, 2] and CDF~[3, 4] collaborations.
It was found~[1--4] that the shape of the measured cross sections as a 
function of photon transverse 
energy $E_T^\gamma$ is poorly described by next-to-leading order (NLO)
QCD calculations: the observed $E_T^\gamma$ distribution is  
steeper than the predictions of perturbative QCD.
These shape differences lead to a significant disagreement
in the ratio of cross sections calculated at different center-of-mass
energies $\sqrt s = 630$ GeV and $\sqrt s = 1800$ GeV as a function of scaling variable
$x_T = 2 E_T^\gamma/\sqrt s$. The disagreement in the
$x_T$ ratio is difficult to explain with conventional
theoretical uncertainties connected with scale dependence and 
parametrizations of the parton distributions~[2, 3].
However, such discrepancy can be reduced~[26, 27] by 
introducing some additional intrinsic transverse momentum $k_T$ of the 
incoming partons, which is usually assumed to have a 
Gaussian-like distribution~[26].
The average value of this $k_T$ increases from 
$\langle k_T \rangle \sim 0.5$ GeV to more than 
$\langle k_T \rangle \sim 2$ GeV as the $\sqrt s$ increases
from UA6 to Tevatron energies~[27].
However, in this case the full kinematics 
of the subprocess is not taken into account, 
as it was argued in Ref.~[25].

The inclusive prompt photon hadroproduction at Tevatron
in the $k_T$-factorization QCD approach was considered in Ref.~[25].
The unintegrated parton distributions in a proton were obtained
using the KMR formalism. The role of the both non-perturbative and 
perturbative components of partonic transverse momentum $k_T$ in describing of 
the observed $E_T^\gamma$ spectrum was investigated. 
However, the KMR unintegrated parton densities were 
obtained in the double leading logarithmic approximation (DLLA) only. 
Also in these calculations the usual on-shell matrix elements 
of hard partonic subprocesses were 
evaluated with precise off-shell kinematics.

In the present paper we apply the KMR method to obtain the unintegrated 
quark and gluon distributions in a proton $f_a(x,{\mathbf k}_T^2,\mu^2)$ 
independently from other authors. Then, we study inclusive 
prompt photon hadroproduction at Tevatron in more detail.
We calculate the double differential cross sections 
$d\sigma/dE_T^\gamma d\eta^\gamma$ at two different center-of-mass 
energies $\sqrt s = 630$ GeV and $\sqrt s = 1800$ GeV and 
compare our theoretical results with the recent D$\oslash$ and CDF 
experimental data~[1--4]. In order to estimate the theoretical uncertainties 
of our predictions we study the renormalization and factorization 
scale dependences of calculated cross sections.
Also we study the ratio of cross sections calculated at 
different center-of-mass energies $\sqrt s = 630$ GeV and $\sqrt s = 1800$ GeV.
After that we extrapolate our predictions to LHC energies.
In all these calculations we use the expressions for the 
partonic off-shell matrix elements which were obtained
in our previous paper~[24].

Also we investigate here the prompt photon and associated muon
production at Fermilab Tevatron. In the Standard Model (SM) these events 
come mainly due to Compton scattering process $g + Q \to \gamma + Q$, 
where final state muon originates from the semileptonic decay of heavy
(charm or bottom) quark $Q$~[5]. It is important that studying of such processes
possible can provide an information about new physics beyond the SM~[28].
Therefore it is necessary to have a realistic estimation of 
associated $\gamma + \mu$ production cross sections within the QCD.
Such calculations in the $k_T$-factorization QCD
approach are performed for the first time.
In order to investigate the underlying dynamics in more detail, we study
the azimuthal correlations between the transverse momenta of 
produced prompt photon and muon. These quantities are sensitive to 
the different production mechanisms and also are powerful tests for 
the non-collinear evolution~[29, 30]. 

The our paper is organized as follows. In Section 2 the KMR unintegrated 
parton densities in a proton are obtained and their properties 
are discussed. 
In particular, we compare the KMR gluon distributions with ones obtained
from the full CCFM equation and within the framework of
the Linked Dipole Chain model~[31] (which is reformulation 
and generalization of the CCFM model).
In Section 3 we present the basic
formulas with a brief review of calculation steps.
In Section 4 we present the numerical results of
our calculations. Finally, in Section 5, we give some conclusions.

\section{The KMR unintegrated partons} \indent 

The Kimber-Martin-Ryskin approach~[22] is the formalism to construct
parton distributions $f_a(x,{\mathbf k}_T^2,\mu^2)$ unintegrated over the parton 
transverse momenta ${\mathbf k}_T^2$ from the known conventional parton
distributions $a(x,\mu^2)$, where $a = xg$ or $a = xq$. This formalism 
is valid for a proton as well as photon and
can embody both DGLAP and BFKL contributions. It also accounts for 
the angular ordering which comes from coherence effects in gluon emission.
The key observation here is that the $\mu$ dependence of the unintegrated 
parton distributions $f_a(x,{\mathbf k}_T^2,\mu^2)$ enters at the last step
of the evolution, and therefore single scale evolution equations (DGLAP 
or unified DGLAP-BFKL~[32]) can be used up to this step. Also it was shown~[22] 
that the unintegrated distributions obtained via unified DGLAP-BFKL 
evolution are rather similar to those based on the pure DGLAP equations.
It is because the imposition of the angular ordering constraint is more 
important than including the BFKL effects. Based on this point, 
in our calculations we use much more simpler DGLAP equation up to 
the last evolution step. In this approximation, the unintegrated quark and 
gluon distributions are given~[22] by
$$
  \displaystyle f_q(x,{\mathbf k}_T^2,\mu^2) = T_q({\mathbf k}_T^2,\mu^2) {\alpha_s({\mathbf k}_T^2)\over 2\pi} \times \atop {
  \displaystyle \times \int\limits_x^1 dz \left[P_{qq}(z) {x\over z} q\left({x\over z},{\mathbf k}_T^2\right) \Theta\left(\Delta - z\right) + P_{qg}(z) {x\over z} g\left({x\over z},{\mathbf k}_T^2\right) \right],} \eqno (1)
$$
$$
  \displaystyle f_g(x,{\mathbf k}_T^2,\mu^2) = T_g({\mathbf k}_T^2,\mu^2) {\alpha_s({\mathbf k}_T^2)\over 2\pi} \times \atop {
  \displaystyle \times \int\limits_x^1 dz \left[\sum_q P_{gq}(z) {x\over z} q\left({x\over z},{\mathbf k}_T^2\right) + P_{gg}(z) {x\over z} g\left({x\over z},{\mathbf k}_T^2\right)\Theta\left(\Delta - z\right) \right],} \eqno (2)
$$

\noindent
where $P_{ab}(z)$ are the usual unregulated leading order DGLAP splitting 
functions, and $q(x,\mu^2)$ and $g(x,\mu^2)$ are the conventional quark 
and gluon densities. The theta functions which appear in (1) and (2) imply 
the angular-ordering constraint $\Delta = \mu/(\mu + |{\mathbf k}_T|)$ 
specifically to the last evolution step to regulate the soft gluon
singularities. For other evolution steps, the strong ordering in 
transverse momentum within the DGLAP equations automatically 
ensures angular ordering. It is important that parton 
distributions $f_a(x,{\mathbf k}_T^2,\mu^2)$ extended now into 
the ${\mathbf k}_T^2 > \mu^2$ region. This fact is in the clear contrast with the 
usual DGLAP evolution\footnote{We would like to note that 
cut-off $\Delta$ can be taken $\Delta = |{\mathbf k}_T|/\mu$ also~[25]. 
In this case the unintegrated parton distributions given by (1) --- (2) 
vanish for ${\mathbf k}_T^2 > \mu^2$ in accordance with 
the DGLAP strong ordering in ${\mathbf k}_T^2$.}.

The virtual (loop) contributions may be resummed 
to all orders by the quark and gluon Sudakov form factors
$$
  \ln T_q({\mathbf k}_T^2,\mu^2) = - \int\limits_{{\mathbf k}_T^2}^{\mu^2} {d {\mathbf p}_T^2\over {\mathbf p}_T^2} {\alpha_s({\mathbf p}_T^2)\over 2\pi} \int\limits_0^{z_{\rm max}} dz P_{qq}(z), \eqno (3)
$$
$$
  \ln T_g({\mathbf k}_T^2,\mu^2) = - \int\limits_{{\mathbf k}_T^2}^{\mu^2} {d {\mathbf p}_T^2\over {\mathbf p}_T^2} {\alpha_s({\mathbf p}_T^2)\over 2\pi} \left[ n_f \int\limits_0^1 dz P_{qg}(z) + \int\limits_{z_{\rm min}}^{z_{\rm max}} dz z P_{gg}(z) \right], \eqno (4)
$$

\noindent
where $z_{\rm max} = 1 - z_{\rm min} = {\mu/({\mu + |{\mathbf p}_T|}})$.
The form factors $T_a({\mathbf k}_T^2,\mu^2)$ give the probability of 
evolving from a scale ${\mathbf k}_T^2$ to a scale $\mu^2$ without 
parton emission. In according with (3) and (4)
$T_a({\mathbf k}_T^2,\mu^2) = 1$ in the ${\mathbf k}_T^2 > \mu^2$ region.

Note that such definition of the $f_a(x,{\mathbf k}_T^2,\mu^2)$ is 
correct for ${\mathbf k}_T^2 > \mu_0^2$ only, where 
$\mu_0 \sim 1$ GeV is the minimum scale for which DGLAP evolution of 
the collinear parton densities is valid. Everywhere in our numerical 
calculations we set the starting scale $\mu_0$ to be equal $\mu_0 = 1$ GeV.
Since the starting point of this derivation is the leading order 
DGLAP equations, the unintegrated parton distributions must satisfy
the normalisation condition
$$
  a(x,\mu^2) = \int\limits_0^{\mu^2} f_a(x,{\mathbf k}_T^2,\mu^2) d{\mathbf k}_T^2. \eqno(5)
$$

\noindent
This relation will be exactly satisfied if we define~[22]
$$
  f_a(x,{\mathbf k}_T^2,\mu^2)\vert_{{\mathbf k}_T^2 < \mu_0^2} = a(x,\mu_0^2) T_a(\mu_0^2,\mu^2). \eqno(6)
$$

\noindent
Then, we have obtained the unintegrated parton distributions in a proton. 
In Figure~1 we show these densities $f_a(x,{\mathbf k}_T^2,\mu^2)$
at scale $\mu^2 = 100\,{\rm GeV}^2$ as a 
function of $x$ for different values of ${\mathbf k}_T^2$, namely 
${\mathbf k}_T^2 = 2\,{\rm GeV}^2$ (a),
${\mathbf k}_T^2 = 5\,{\rm GeV}^2$ (b),
${\mathbf k}_T^2 = 10\,{\rm GeV}^2$ (c) and 
${\mathbf k}_T^2 = 20\,{\rm GeV}^2$ (d). 
The solid, dashed, short dashed, dotted, dash-dotted and short dash-dotted lines
correspond to the unintegrated gluon (divided by factor $10$), 
$u + \bar u$, $d + \bar d$, $s$, $c$ and $b$ quark distributions, 
respectively. We have used here the standard leading-order 
Gl\"uck-Reya-Vogt (GRV) parametrizations~[33] of the collinear quark 
and gluon densities $a(x,\mu^2)$. 
Note that unintegrated $c$ and $b$ quark distributions at 
${\mathbf k}_T^2 = 2\,{\rm GeV}^2$ are very similar and cannot
be really resolved in Figure~1.
Also we have checked numerically that normalization condition (5) is correctly 
satisfied for all unintegrated 
parton distributions $f_a(x,{\mathbf k}_T^2,\mu^2)$.

It is interesting to compare the KMR-constructed unintegrated parton densities  
with the distributions obtained in other approaches. 
Recently the full CCFM equation in a proton was solved numerically 
using a Monte Carlo method, and new fits of the unintegrated gluon 
distributions (J2003 set 1 --- 3) have been presented~[34].
The input parameters were fitted to describe the proton structure 
function $F_2(x,Q^2)$. These unintegrated gluon densities were used
also in analysis of the forward jet production at HERA, 
charm and bottom production 
at Tevatron~[34], and charm and $J/\psi$ production at 
LEP2 energies~[35].
Also three different versions of the unintegrated gluon 
distributions obtained in the framework of the 
Linked Dipole Chain (LDC) model
have been presented~[36]. These gluon densities has 
been fitted to the inclusive $F_2$ data at HERA and already
were used, in particular, in analysis~[37] of the charm and beauty hadroproduction 
at Tevatron.
In Figure~2 we plot the KMR (as a solid lines), the J2003 set 1 (as a dashed lines) 
unintegrated gluon distributions and so called
{\it standard} version of the LDC unintegrated gluon distribution
(as a dash-dotted lines) at scale $\mu^2 = 100\,{\rm GeV}^2$ 
as a function of $x$ for different 
values of gluon ${\mathbf k}_T^2$, namely 
${\mathbf k}_T^2 = 2\,{\rm GeV}^2$ (a),
${\mathbf k}_T^2 = 10\,{\rm GeV}^2$ (b),
${\mathbf k}_T^2 = 20\,{\rm GeV}^2$ (c) and 
${\mathbf k}_T^2 = 50\,{\rm GeV}^2$ (d). 
One can see that behaviour of different unintegrated
parton distributions in the small ${\mathbf k}_T^2$ region (which essentially drives
the total cross sections) is very different.
At the same time the differences between these distributions tends to be 
small when gluon transverse momentum ${\mathbf k}_T^2$ is large.
Therefore the dependence of our predictions on the evolution scheme 
possible may be rather large, and further theoretical attempts are necessary 
to reduce this uncertainty.

\section{Calculation details} 

\subsection{The subprocesses under consideration} \indent 

The main contribution to the prompt photon production in 
proton-antiproton collisions at Tevatron and LHC colliders 
comes from quark-gluon and quark-antiquark induced partonic subprocesses:
$$
  q(k_1) + g(k_2) \to \gamma (p^\gamma) + q (p^{\prime}), \eqno(7)
$$
$$
  q(k_1) + \bar q (k_2) \to \gamma (p^\gamma) + g (p^{\prime}), \eqno(8)
$$

\noindent 
where the particles four-momenta are given in parentheses.
Additionally, photons can be produced 
through the fragmentation of a partonic jet into a single photon 
carrying a large fraction $z$ of the jet energy~[7]. These processes are 
described in terms of quark-to-photon $D_{q\to\gamma}(z,\mu^2)$ and
gluon-to-photon $D_{g\to\gamma}(z,\mu^2)$ fragmentation functions~[8]. 
The main feature of the fragmentation contribution in leading order
is fact that produced jet is balanced by a photon on the opposite side of
the event and accompanied by collinear photon on the same side of the event.

It is important that in order to reduce the huge background
from the secondary photons produced by the decays of $\pi^0$ and $\eta$ 
mesons the isolation criterion is introduced in the experimental analyses.
This criterion is the following. A photon is isolated if the 
amount of hadronic transverse energy $E_T^{\rm had}$, deposited inside
a cone with aperture $R$ centered around the photon direction in the 
pseudo-rapidity and azimuthal angle plane, is smaller than
some value $E_T^{\rm max}$:
$$
  \displaystyle E_T^{\rm had} \le E_T^{\rm max},\atop
  \displaystyle (\eta - \eta^{\gamma})^2 + (\phi - \phi^{\gamma})^2 \le R^2. \eqno(9)
$$

\noindent 
The both D$\oslash$ and CDF collaborations take $R \sim 0.4$ and 
$E_T^{\rm max} \sim 1$ GeV in the experiment~[1--5]. 
Isolation not only reduces the background 
but also significantly reduces the fragmentation components.
It was shown~[38] that after applying the isolation cut (9) the contribution from the
fragmentation subprocesses is about 10\% of the total cross section.
Since the dependence of our results on the non-collinear parton evolution
scheme may be rather large (as it was demonstrated in Section~2), 
in our further analysis we will neglect the relative small 
fragmentation contribution and consider only the 
$q + g \to \gamma + q$ and $q + \bar q \to \gamma + g$ subprocesses. 
We would like to note that photon produced in 
these processes is isolated from the quark or gluon jet by 
requiring a non-zero transverse momentum of a photon or jet in 
the $p\bar p$ center-of-mass frame.

\subsection{Kinematics} \indent 

Let $p_1$ and $p_2$ be the four-momenta 
of the incoming protons. The initial off-shell partons have the four-momenta
$k_1$ and $k_2$. In our analysis below we will use the Sudakov 
decomposition, which has the following form:
$$
  p^{\gamma} = \alpha_1 p_1 + \beta_1 p_2 + p^{\gamma}_{T},\quad p^\prime = \alpha_2 p_1 + \beta_2 p_2 + p_T^\prime,\atop
  k_1 = x_1 p_1 + k_{1T},\quad k_2 = x_2 p_2 + k_{2T}, \eqno(7)
$$

\noindent 
where $k_{1T}$, $k_{2T}$, $p_T^\gamma$ and $p_T^\prime$ are the
transverse four-momenta of the corresponding particles.
It is important that ${\mathbf k}_{1T}^2 = - k_{1T}^2 \neq 0$ and
${\mathbf k}_{2T}^2 = - k_{2T}^2 \neq 0$. In the $p \bar p$ center-of-mass 
frame we can write
$$
  p_1 = {\sqrt s}/2 (1,0,0,1),\quad p_2 = {\sqrt s}/2 (1,0,0,-1), \eqno(8)
$$

\noindent
where $s = (p_1 + p_2)^2$ is the total energy of the process under consideration
and we neglect the masses of the incoming protons. The Sudakov variables
are expressed as follows:
$$
  \displaystyle \alpha_1={E_T^\gamma\over {\sqrt s}}\exp(y^\gamma),\quad \alpha_2={m_T^\prime\over {\sqrt s}}\exp(y^\prime),\atop
  \displaystyle \beta_1={E_T^\gamma\over {\sqrt s}}\exp(-y^\gamma),\quad \beta_2={m_T^\prime\over {\sqrt s}}\exp(-y^\prime), \eqno(9)
$$

\noindent
where $E_T^\gamma$ and $m_T^\prime$ are the transverse enegy and transverse mass 
of produced photon and outgoing parton, respectively, 
and $y^\gamma$ and $y^\prime$ 
are their rapidities (in the $p\bar p$ center-of-mass frame). 
The photon pseudo-rapidity $\eta^\gamma$ is defined as 
$\eta^\gamma = - \ln \tan (\theta^\gamma/2)$,
where $\theta^\gamma$ is the polar angle of the prompt photon 
with respect to the proton beam.
From the conservation laws we can easily obtain the following conditions:
$$
  x_1 = \alpha_1 + \alpha_2,\quad x_2 = \beta_1 + \beta_2,\quad {\mathbf k}_{1T} + {\mathbf k}_{2T} = {\mathbf p}_T^\gamma + {\mathbf p}_{T}^\prime. \eqno(10)
$$

\noindent
The scaling variable $x_T = 2 E_T^\gamma/\sqrt s$ is also often used
in analysis of the prompt photon production.

\subsection{Cross section for prompt photon production} \indent 

The total cross section for prompt photon hadroproduction
at high energies in the $k_T$-factorization QCD approach can be written as
$$
  \displaystyle d\sigma (p + \bar p \to \gamma + X) = \sum_{a,b} \int {dx_1\over x_1} f_a(x_1,{\mathbf k}_{1 T}^2,\mu^2) d{\mathbf k}_{1 T}^2 {d\phi_1\over 2\pi} \times \atop 
  \displaystyle \times \int {dx_2\over x_2} f_b(x_2,{\mathbf k}_{2 T}^2,\mu^2) d{\mathbf k}_{2 T}^2 {d\phi_2\over 2\pi} d{\hat \sigma} (a b \to \gamma c), \eqno(11)
$$

\noindent
where $a \ldots c = q$ and/or $g$, ${\hat \sigma} (a b \to \gamma c)$ is the 
cross section of the photon production in the corresponding 
parton-parton interaction. Here initial partons $a$ and $b$ have 
longitudinal momentum fractions $x_1$ and $x_2$, non-zero 
transverse momenta ${\mathbf k}_{1 T}$ and ${\mathbf k}_{2 T}$
and azimuthal angles $\phi_1$ and $\phi_2$.
From the expression (11) we can easily obtain the formula:
$$
  \displaystyle \sigma(p + \bar p \to \gamma + X) = \sum_{a,b} \int {E_T^\gamma\over 8\pi (x_1 x_2 s)^2} |\bar {\cal M}|^2(a b \to \gamma c) \times \atop 
  \displaystyle \times f_a(x_1,{\mathbf k}_{1 T}^2,\mu^2) f_b(x_2,{\mathbf k}_{2 T}^2,\mu^2) d{\mathbf k}_{1 T}^2 d{\mathbf k}_{2 T}^2 dE_T^\gamma dy^\gamma dy^c {d\phi_1\over 2\pi} {d\phi_2\over 2\pi} {d\phi^\gamma\over 2\pi}, \eqno(12)
$$

\noindent
where $|\bar {\cal M}|^2 (a b \to \gamma c)$ is the hard partonic squared off-mass shell 
matrix element which depends on the transverse momenta ${\mathbf k}_{1 T}^2$ and 
${\mathbf k}_{2 T}^2$, $y^c$ is the rapidity of the parton $c$ in the 
$p\bar p$ center-of-mass frame, $\phi_1$, $\phi_2$ and $\phi^\gamma$ are the 
azimuthal angles of the initial partons $a$, $b$ and produced prompt photon,
respectively. The analytic expression for the $|\bar {\cal M}|^2 (a b \to \gamma c)$
was obtained in our previous paper~[24].
We would like to note that if we average the expression (12) over 
${\mathbf k}_{1 T}$ and ${\mathbf k}_{2 T}$ and take the limit 
${\mathbf k}_{1 T}^2 \to 0$ and ${\mathbf k}_{2 T}^2 \to 0$,
then we obtain well-known expression for the prompt photon 
hadroproduction in leading-order (LO) perturbative QCD.

The multidimensional integration in (12) has been performed
by means of the Monte Carlo technique, using the routine VEGAS~[39].
The full C$++$ code is available from the authors on 
request\footnote{lipatov@theory.sinp.msu.ru}.

\section{Numerical results} \indent 

We now are in a position to present our numerical results. First we describe our
theoretical input and the kinematical conditions. After we fixed the unintegrated
parton distributions in a proton $f_a(x,{\mathbf k}_{T}^2,\mu^2)$, 
the cross section (12) depends on the energy scale $\mu$. 
As it often done~[38] for prompt photon production, we choose the factorization and 
renormalization scales to be equal $\mu_F = \mu_R = \mu = \xi E_T^\gamma$. 
In order to estimate the theoretical uncertainties of our calculations
we will vary the scale parameter $\xi$ between 1/2 and 2 about the default 
value $\xi = 1$. Also we use LO formula
for the strong coupling constant $\alpha_s(\mu^2)$ with $n_f = 3$ active 
(massless) quark flavours and $\Lambda_{\rm QCD} = 232$ MeV, 
such that $\alpha_s(M_Z^2) = 0.1169$.
In our analysis we not neglect the 
charm and bottom quark masses and set them to be 
$m_c = 1.5$ GeV and $m_b = 4.75$ GeV, respectively.

\subsection{Inclusive prompt photon production at Tevatron} \indent 

Experimental data~[1--4] for the inclusive prompt photon 
hadroproduction $p + \bar p \to \gamma + X$ come from both the
D$\oslash$ and CDF collaborations. The D$\oslash$~[1, 2] data 
were obtained in the central and forward pseudo-rapidity
regions for two different center-of-mass energies, namely 
$\sqrt s = 630$ GeV and $\sqrt s = 1800$ GeV.
The central pseudo-rapidity region is defined by the requirement
$|\eta^\gamma| < 0.9$, and the forward one is 
defined by $1.6 < |\eta^\gamma| < 2.5$.
The more recent CDF data~[3] refer to the same central kinematical region
$|\eta^\gamma| < 0.9$ 
for both beam energies $\sqrt s = 630$ GeV and $\sqrt s = 1800$ GeV.
Also very recently the CDF collaboration has presented a new 
measurement~[4] of the prompt photon cross section at $\sqrt s = 1800$ GeV.
This measurement is based on events where the photon converts
into an electron-positron pair in the material of inner detector,
resulting in a two track event signature ("conversion" data).
These data refer only to the central kinematical region. 
Actually, there are two different datasets, 
which were used in the CDF measurement with conversions,
namely 8 GeV electron data and 23 GeV photon 
data\footnote{See Ref.~[4] for more details.}.
In all these measurements the double differential
cross sections $d\sigma/dE_T^\gamma d \eta^\gamma$
as a function of the transverse energy $E_T^\gamma$ 
are determined.

The results of our calculations are shown in Figs.~3 --- 9. 
So, Figs.~3 and~4 confront the cross sections 
$d\sigma/dE_T^\gamma d \eta^\gamma$ calculated at $\sqrt s = 630$ GeV 
in different kinematical regions with the D$\oslash$~[2] and CDF~[3] data.
The solid lines are obtained by fixing both 
the factorization and 
renormalization scales at the default value $\mu = E_T^\gamma$
whereas upper and lower dashed lines 
correspond to the $\mu = E_T^\gamma/2$ and $\mu = 2 E_T^\gamma$ scales, 
respectively. One can see that our predictions agree  
with the experimental data within the scale uncertainties.
However, the results of calculation with the default scale 
tend to underestimate the data in the central kinematical region 
and agree with the D$\oslash$ data in the forward $\eta^\gamma$ region. 
The collinear NLO QCD calculations~[38] give 
the similar description of the data: the results of measurement are higher than 
the NLO prediction at low $E_T^\gamma$ in the central $\eta^\gamma$ range 
but agree at all $E_T^\gamma$ in the forward pseudo-rapidity region.
Then, one can see that the scale dependence of our results is
rather large: the variation in $\mu$ as it was described above
changes the cross sections by about 20 --- 30\%. 
The theoretical uncertainties of the collinear NLO
calculations are similar (about 20\%)~[2]. 
However, one should keep in mind
that additional dependence of our results on the evolution scheme may be 
also rather large (as it was discussed in Section~2), 
and overall agreement with the experimental data 
can be improved when unintegrated quark and gluon distributions in a proton
will be studied more detail. At the same time the use of different sets of the 
parton distributions in NLO calculations changes the cross
sections by less than 5\%~[2, 3].

The double differential cross sections $d\sigma/dE_T^\gamma d \eta^\gamma$ 
compared with the experimental data at $\sqrt s = 1800$ GeV 
in different pseudo-rapidity regions are
shown in Figs.~5 --- 7. All curves here are the same as in Fig.~3.
We find that our predictions agree well with the 
D$\oslash$~[1] and CDF~[3, 4] data both in normalization and shape. 
There are only rather 
small overestimation of the data at low $E_T^\gamma$ values in
Figs.~6 and~7. Again, the scale dependence of our calculations 
is about 20 --- 30\%. 
The theoretical uncertainties of the collinear NLO predictions are 
much smaller, about 6\%~[1]. 
Note that now NLO calculations
agree with the data more qualitatively. 
So, the shape of the measured cross sections is generally 
steeper than that of the NLO predictions. It was shown~[3, 4] that this shape 
difference is difficult to explain simply by changing the 
renormalization/factorization scales in the collinear calculation, 
or the set of parton distribution functions.

Also the disagreement between data and NLO calculations is  
visible~[2, 3] in the ratio of the cross sections 
at different energies. This quantity is known 
as a very informative subject of investigations and 
provides a precise test of the QCD calculations. It is because 
many factors which affect the absolute normalization, as well as many 
theoretical and experimental uncertainties partially or completely 
cancel out~[2, 3]. In particular, the cross section ratio
provides a direct probe of the matrix elements of the 
hard partonic subprocesses since the theoretical uncertainties due to 
the quark and gluon distributions are reduced.

So, the D$\oslash$ collaboration has published the results of measurement~[2] 
for the ratio of 630 GeV and 1800 GeV dimensionless cross sections
$\sigma_D$ as a function of scaling variable $x_T$. The 
measured cross section $\sigma_D$ averaged over azimuth is defined as
$$
  \sigma_D = {1\over 2\pi} (E_T^\gamma)^3 {d \sigma \over d E_T^\gamma d \eta^\gamma}. \eqno(13)
$$

\noindent 
The ratio $\sigma_D(630\,{\rm GeV})/\sigma_D(1800\,{\rm GeV})$ 
compared with the D$\oslash$ experimental data~[2] 
in different pseudo-rapidity $\eta^\gamma$ regions is shown in Figs.~8 and~9.
The solid lines represent the $k_T$-factorization 
predictions at default scale $\mu$. For comparison we show also the results 
of the collinear leading-order (LO) QCD calculations with 
the GRV parton densities~[33] of a proton (as a dashed lines). 
Note that when we perform the LO QCD calculations 
we take into account the partonic subprocesses (7) and (8)
and neglect the small fragmentation contributions, as it was done in the
$k_T$-factorization case. It is clear that although the experimental 
points have large errors they tend to support the $k_T$-factorization
predictions. We would like to point out again that now sensitivity 
of our results to the non-collinear evolution scheme is minimized.
In the collinear approach, the NLO corrections improve the description of the data
and then sum of LO and NLO contributions practically coincides with our
results at $x_T > 0.05$~[2]. This fact is clear indicates 
that the main part of the collinear high-order corrections is 
already included at leading-order level in the $k_T$-factorization
formalism\footnote{See Refs.~[15, 16] for more details.}.
Nevertheless, the experimental data at the lowest $x_T$ are 
systematically higher~[2] than NLO QCD predictions in both central 
and forward pseudo-rapidity regions, and it was claimed~[3] that
ratio of the two cross sections as a function of the $x_T$
is difficult to reconcile with the NLO QCD calculations.

\subsection{Associated prompt photon and muon production at Tevatron} \indent 

Now we investigate the prompt photon and associated muon production 
at Tevatron. These events are assumed to come from
associated prompt photon and heavy (charm or bottom) quark production with 
the heavy quark decaying into a muon~[5]. To calculate muon production
from heavy quarks we first 
convert charm (bottom) quarks into $D$ ($B$) hadrons using
the Peterson fragmentation function~[40] and then 
simulate their semileptonic decay according to  
Standard Model. Our default set 
of the fragmentation parameters and branching rations
is $\epsilon_c = 0.06$, $f(c \to \mu) = 10.2$\% and 
$\epsilon_b = 0.06$, $f(b \to \mu) = 10.8$\%.
Of course, muon transverse momenta spectra
are sensitive to the fragmentation functions. 
However, this dependence is expected to be small as compared 
with the scale uncertainties and the uncertainties connected with  
unintegrated parton densities. On the other hand, the
variations of $\epsilon_c$ and $\epsilon_b$ do not affect on the azimuthal 
angle distribution (which is one of the main subject of our 
study) because a fragmentation does not change
the direction of the quark or hadron momentum.

The experimental data for the $\gamma + \mu$ cross section at 
Tevatron come from CDF collaboration~[5]. 
The differential cross section $d\sigma/d p_T^\gamma$ at $\sqrt s = 1800$ GeV
was obtained.
The photon pseudo-rapidity is required to be within 
$|\eta^\gamma| < 0.9$ region whereas the muon transverse momentum and 
pseudo-rapidity are required to be $p_T^\mu > 4$ GeV and 
$|\eta^\mu| < 1.0$.

The transverse momentum distribution $d\sigma/d p_T^\gamma$
in comparison to experimental data~[5] is shown in Fig.~10.
All curves here are the same as in Fig.~3.
One can see that the shape of this distribution
is well described by our calculations. However, the
theoretical results slightly overestimate the data
in absolute normalization. 
This fact can be connected with the GRV~[33] parametrization
of charm and bottom collinear densities of a proton 
(which are one of the basic ingredients 
of our derivation). Note, however, that in general 
the experimental points still lie within scale uncertainties (about 30\%)
of our calculations. It is important also that our 
predictions practically coincide with the results of 
collinear NLO QCD calculations~[41], 
which are much larger than LO ones~[5]. This fact clearly demonstrates again 
that the main part of the standard high-order corrections is 
already effectively included in the $k_T$-factorization
approach.

Further understanding of the process dynamics 
and in particular of the high-order effects 
may be obtained from the angular correlation between the
transverse momenta of the final state particles.
It was shown~[30] that investigation of these correlations
is a powerful test for the non-collinear parton evolution dynamics.
This is because such quantities are sensitive to
relative contributions of different production
mechanisms to the total cross section.
So, in the collinear LO approximation, the prompt 
photon and heavy quark $Q$ are produced back-to-back. Therefore distribution
over the azimuthal angle difference $\Delta\phi^{\gamma Q}$ must
be simply a delta function at $\Delta\phi^{\gamma Q} = \pi$.
Taking into account the non-vanishing initial parton
transverse momenta ${\mathbf k}_{1 T}$ and ${\mathbf k}_{2 T}$ leads 
to the violation of this back-to-back kinematics in the 
$k_T$-factorization approach.

The differential cross section $d\sigma/d\Delta\phi^{\gamma \mu}$
calculated at $p_T^\mu > 4$ GeV, $|\eta^\mu| < 1.0$ and $|\eta^\gamma| < 0.9$
is shown in Fig.~11. The solid and both dashed lines
here are the same as in Fig.~3. The LO QCD contribution 
also shown (as a dash-dotted line).
One can see a clear difference in shape between 
$k_T$-factorization results and collinear LO QCD ones. 
As it was expected, the LO QCD predicts only a 
peak at $\Delta\phi^{\gamma \mu} = \pi$. 
The small broadering of $\Delta\phi^{\gamma \mu}$ distribution
illustrates the effect of $Q \to \mu$ decays.
Unfortunately, the predictions of the NLO QCD for this
distribution are still unknown. 
The direct comparison between NLO calculations and our results
should give a number of interesting insighs.
In particular, it can provide an important information about high-order terms
which are missed in the leading-order $k_T$-factorization approach.
In any case, the future theoretical and experimental study of such processes
will be important check of non-collinear parton evolution.

\subsection{Inclusive prompt photon production at LHC} \indent 

We can conclude from Figs.~3 --- 11 that our calculations in general
agree well with experimental data~[1--5] taken by the D$\oslash$ and CDF
collaborations at Tevatron. Based on this point,
now we can try to extrapolate our theoretical predictions
to LHC energies. We perform the calculation for both 
central and forward pseudo-rapidities $\eta^\gamma$.
As a representative example,
we will define the central and forward kinematical regions 
by the requirements $|\eta^\gamma| < 2.5$ and 
$2.5 < |\eta^\gamma| < 4$, respectively.

The transverse energy $E_T^\gamma$
distributions of the inclusive prompt photon production in different 
$\eta^\gamma$ ranges at $\sqrt s = 14$ TeV
are shown in Figs.~12 and~13. All curves here are the same as 
in Fig.~3. One can see that variation in scale $\mu$
changes the estimated cross sections by about 20 --- 30\%. 
However, as it was already discussed above, there are 
additional theoretical uncertainties due to 
the non-collinear parton evolution, and these uncertainties are not 
well studied up to this time. Also the extrapolation
of the available parton distribution to the region
of lower $x$ is a special problem at the LHC 
energies. In particular, one of the problem
is connected with the correct treatment of saturation effects
in small $x$ region\footnote{See also Ref.~[16] for more information}. 
Therefore much more work needs to be done
until these uncertainties will be reduced.

\section{Conclusions} \indent 

We present calculations of the prompt photon hadroproduction 
at high energies 
in the $k_T$-factorization QCD approach. In order to obtain the unintegrated 
quark and gluon distributions in a proton we have used the Kimber-Martin-Ryskin 
prescription. We have investigated both inclusive prompt photon 
and associated with muon production rates. The associated $\gamma + \mu$
events come primarily due to the Compton scattering process $g + Q \to \gamma + Q$, 
with the final state heavy (charm or bottom) quark $Q$ producing a muon. 
The calculations of such cross sections in the $k_T$-factorization 
approach were performed for the first time.

We have found that our predictions for the inclusive prompt photon production 
agree well with experimental data taken by the D$\oslash$ and CDF 
collaborations at Tevatron in both central and forward 
pseudo-rapidity regions. It is very important that perfect 
agreement was found also in the ratio of two cross sections calculated 
at $\sqrt s = 630$ GeV and $\sqrt s = 1800$ GeV. This ratio
provides a direct probe of the QCD matrix elements 
since in this case the theoretical uncertainties connected to the parton
distributions are significantly reduced. We have also demonstrated that the
main part of the standard high-order corrections is already included in 
the $k_T$-factorization formalism at LO level. Additionally, we 
present our predictions for the inclusive prompt photon
production at LHC.

At the same time our results for associated $\gamma + \mu$ production 
tend to overestimate the CDF data but still agree
with data within the scale uncertainties. 
We have demonstrated also the significant deviation
from back-to-back kinematics in prompt 
photon and associated muon production. We can expect that further theoretical and 
experimental study of such processes will give an important information 
about non-collinear parton evolution dynamics.

In order to investigate the theoretical uncertainties of our 
results we have studied the sensitivity of our predictions to the 
renormalization and factorization scales. We have found that 
this dependence is about 20 --- 30\% in wide center-of-mass energy 
range. There are, of course, also additional uncertainties due 
to unintegrated parton distributions in a proton. Therefore much 
more work needs to be done before these uncertainties 
will be reduced. Finally, in our analysis we neglect 
the contribution from the fragmentation processes.
We plan to investigate these problems in more detail
in the forthcoming publications. 

\section{Acknowledgements} \indent 

The authors are very grateful to S.P.~Baranov for encouraging interest
and helpful discussions. This research was supported in part by the 
FASI of Russian Federation (grant NS-1685.2003.2).

\newpage

\begin{figure}
\epsfig{figure=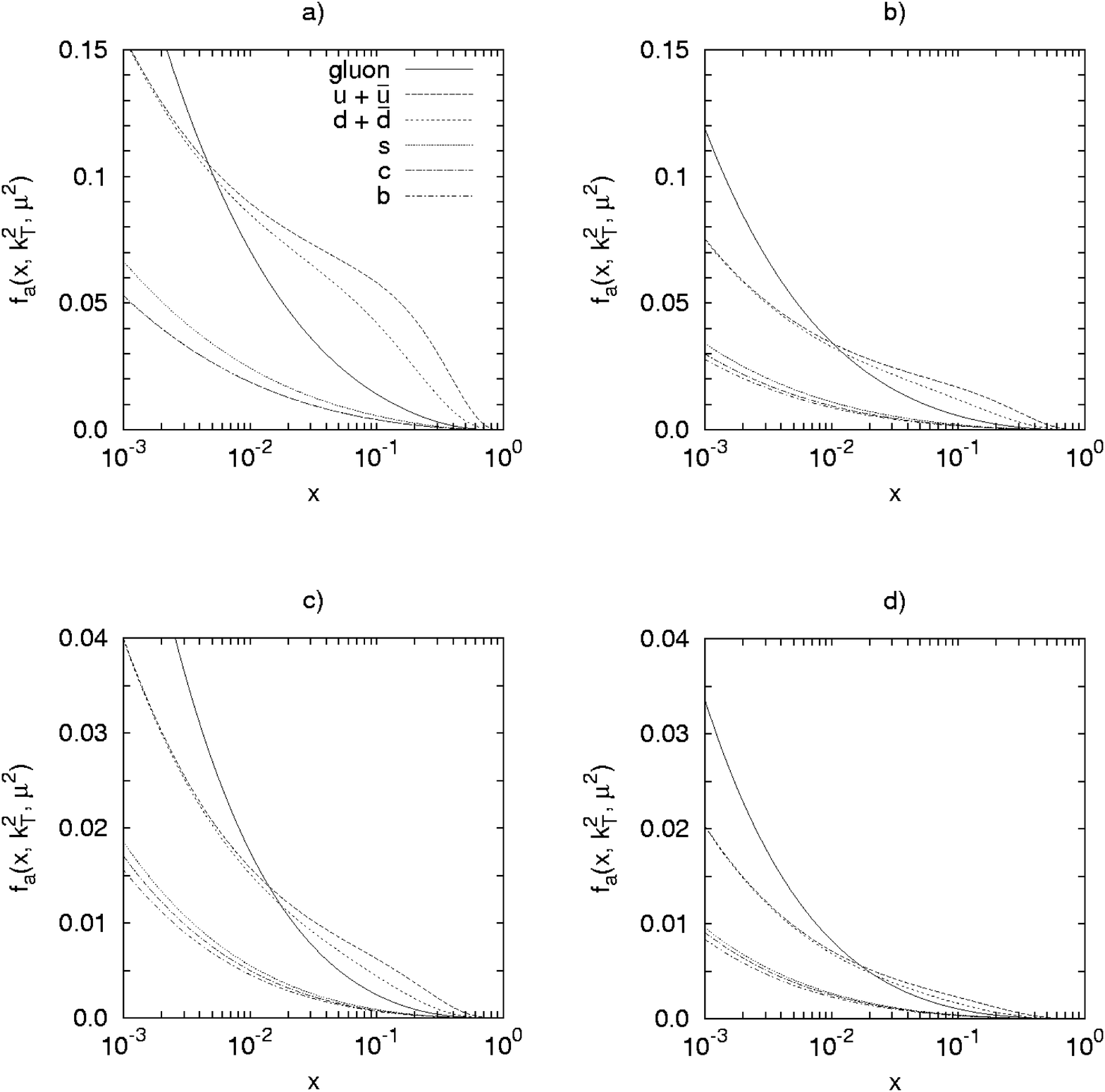, width = 17.5cm, height = 17.5cm}
\caption{The unintegrated parton distributions $f_a(x,{\mathbf k}_T^2,\mu^2)$
at scale $\mu^2 = 100\,{\rm GeV}^2$ as a function 
of $x$ for different values of ${\mathbf k}_T^2$, namely 
${\mathbf k}_T^2 = 2\,{\rm GeV}^2$ (a),
${\mathbf k}_T^2 = 5\,{\rm GeV}^2$ (b),
${\mathbf k}_T^2 = 10\,{\rm GeV}^2$ (c) and 
${\mathbf k}_T^2 = 20\,{\rm GeV}^2$ (d).
The solid, dashed, short dashed, dotted, dash-dotted and short dash-dotted lines
correspond to the unintegrated gluon (divided by factor $10$), 
$u + \bar u$, $d + \bar d$, $s$, $c$ and $b$ quark distributions, respectively.}
\label{fig1}
\end{figure}

\newpage

\begin{figure}
\epsfig{figure=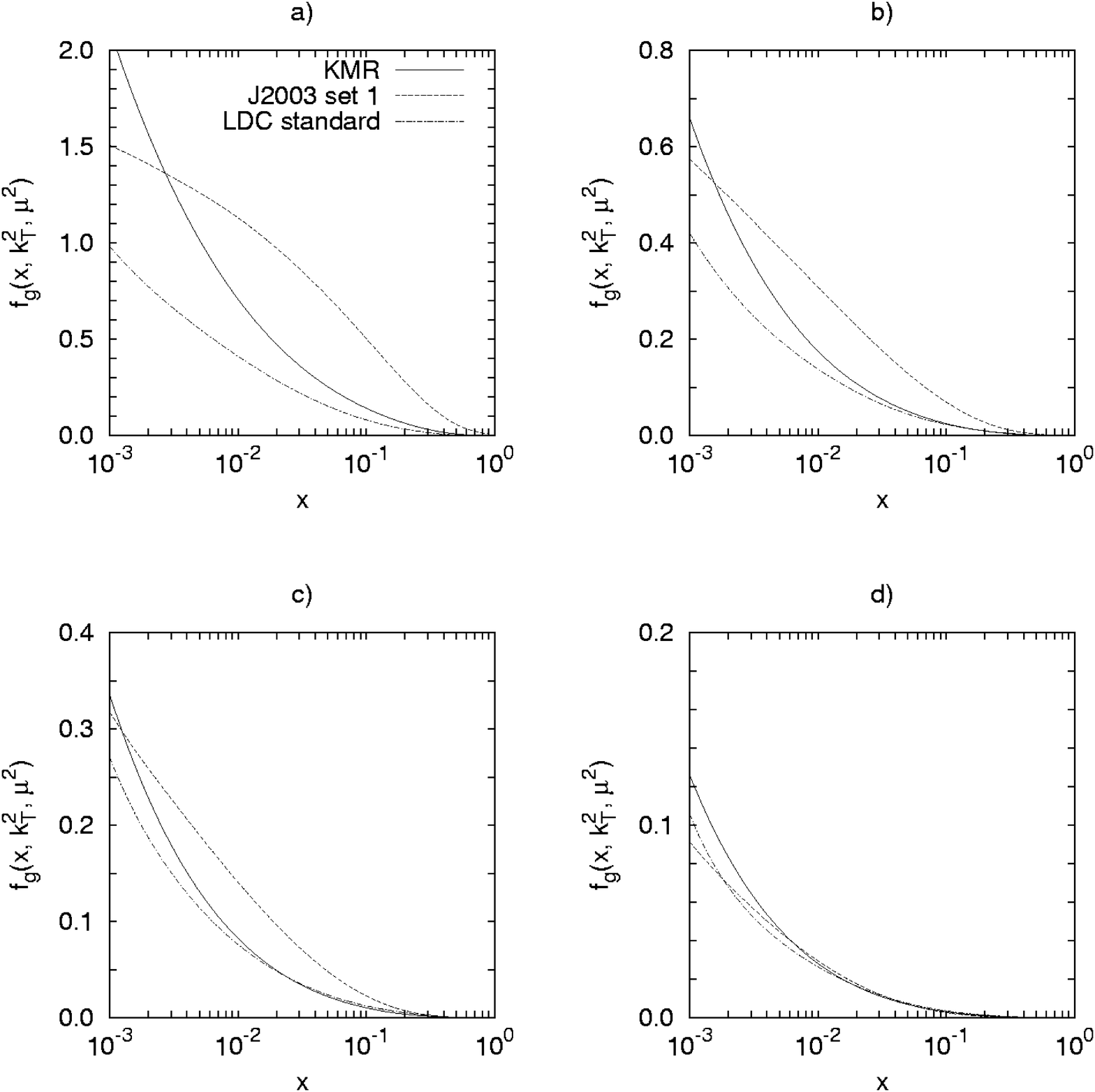, width = 17.5cm, height = 17.5cm}
\caption{The unitegrated gluon distributions $f_g(x,{\mathbf k}_T^2,\mu^2)$
at scale $\mu^2 = 100\,{\rm GeV}^2$ as a function 
of $x$ for different values of ${\mathbf k}_T^2$, namely 
${\mathbf k}_T^2 = 2\,{\rm GeV}^2$ (a),
${\mathbf k}_T^2 = 10\,{\rm GeV}^2$ (b),
${\mathbf k}_T^2 = 20\,{\rm GeV}^2$ (c) and 
${\mathbf k}_T^2 = 50\,{\rm GeV}^2$ (d).
The solid, dashed and dash-dotted lines correspond to the KMR, J2003 set 1 and 
{\it standard} version of the LDC unintegrated gluon densities, respectively.}
\label{fig2}
\end{figure}

\newpage

\begin{figure}
\epsfig{figure=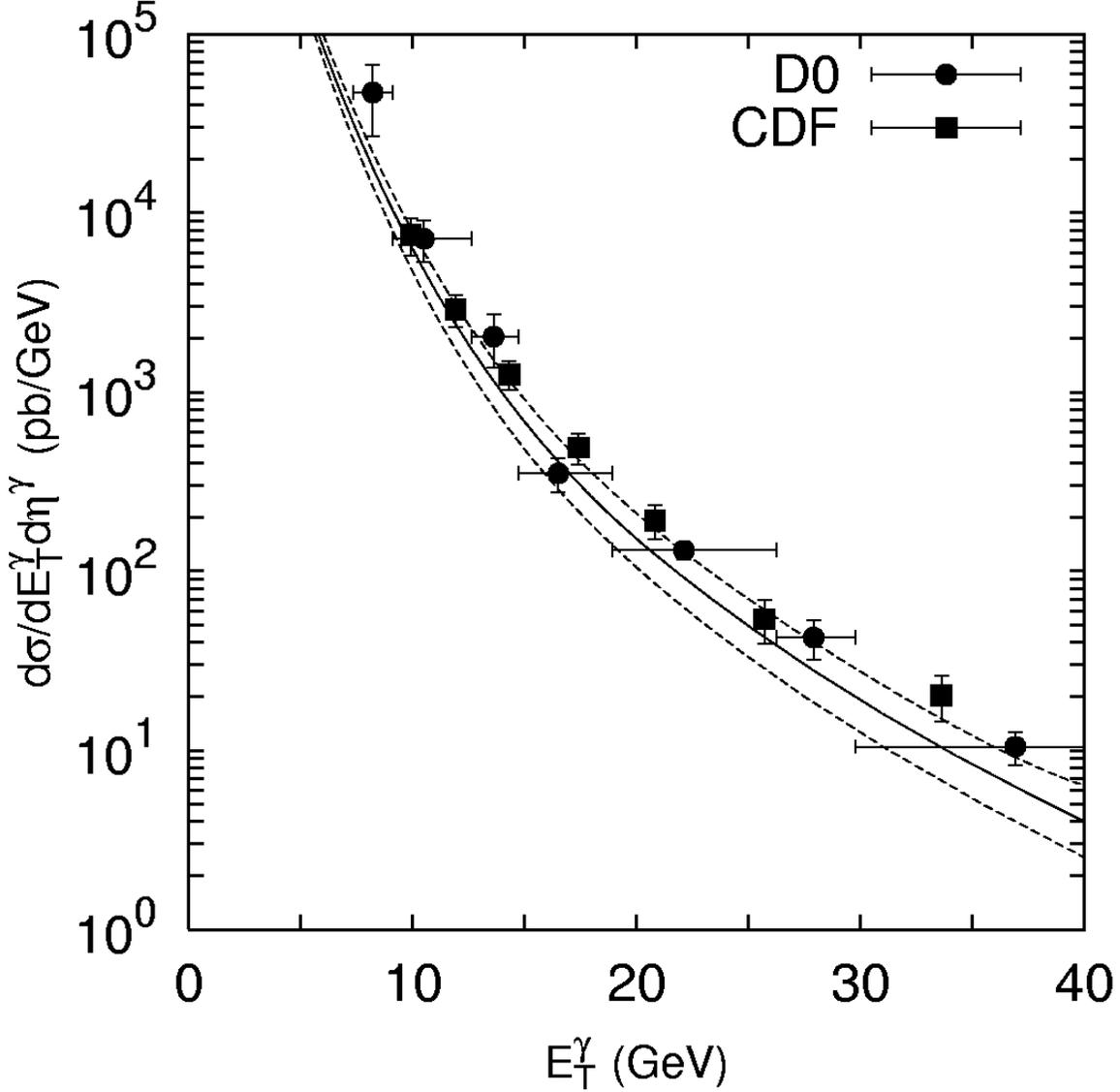, width = 22cm}
\caption{The double differential cross section $d\sigma/d E_T^\gamma d\eta^\gamma$
for inclusive prompt photon hadroproduction 
at $|\eta^\gamma| < 0.9$ 
and $\sqrt s = 630$ GeV. The solid line corresponds to the default 
scale $\mu = E_T^\gamma$, whereas upper and lower dashed lines
correspond to the $\mu = E_T^\gamma/2$ and $\mu = 2 E_T^\gamma$ 
scales, respectively. The experimental data are from D$\oslash$~[2] and CDF~[3].}
\label{fig3}
\end{figure}

\newpage

\begin{figure}
\epsfig{figure=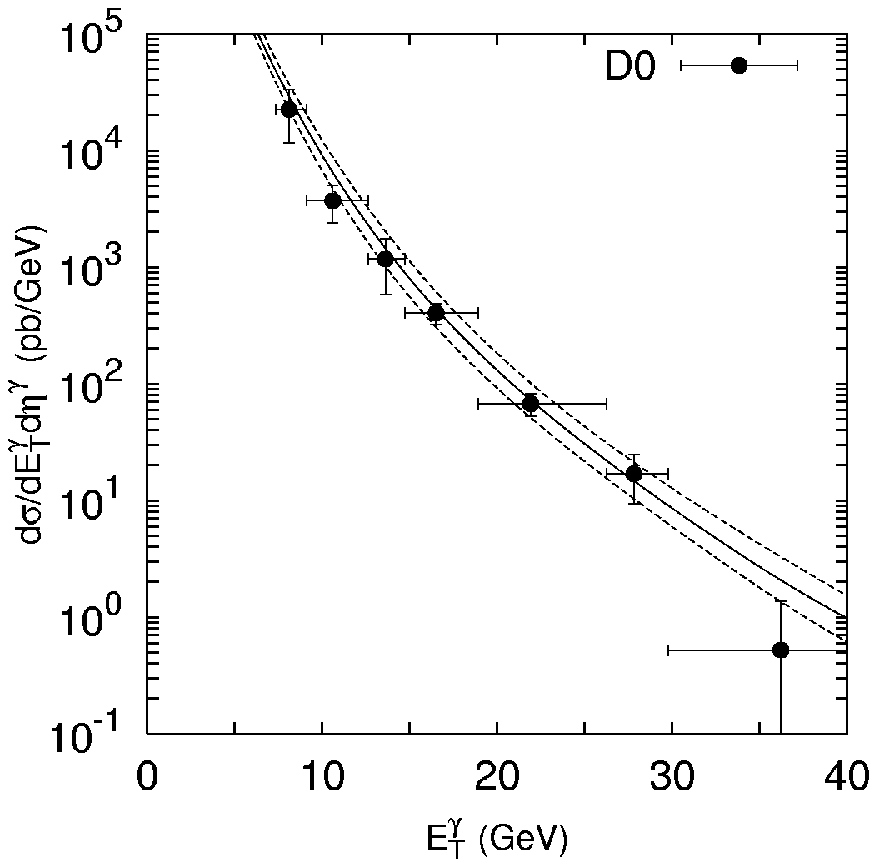, width = 22cm}
\caption{The double differential cross section $d\sigma/d E_T^\gamma d\eta^\gamma$
for inclusive prompt photon hadroproduction 
at $1.6 < |\eta^\gamma| < 2.5$ 
and $\sqrt s = 630$ GeV. All curves are the same as in Figure~3. 
The experimental data are from D$\oslash$~[2].}
\label{fig4}
\end{figure}

\newpage

\begin{figure}
\epsfig{figure=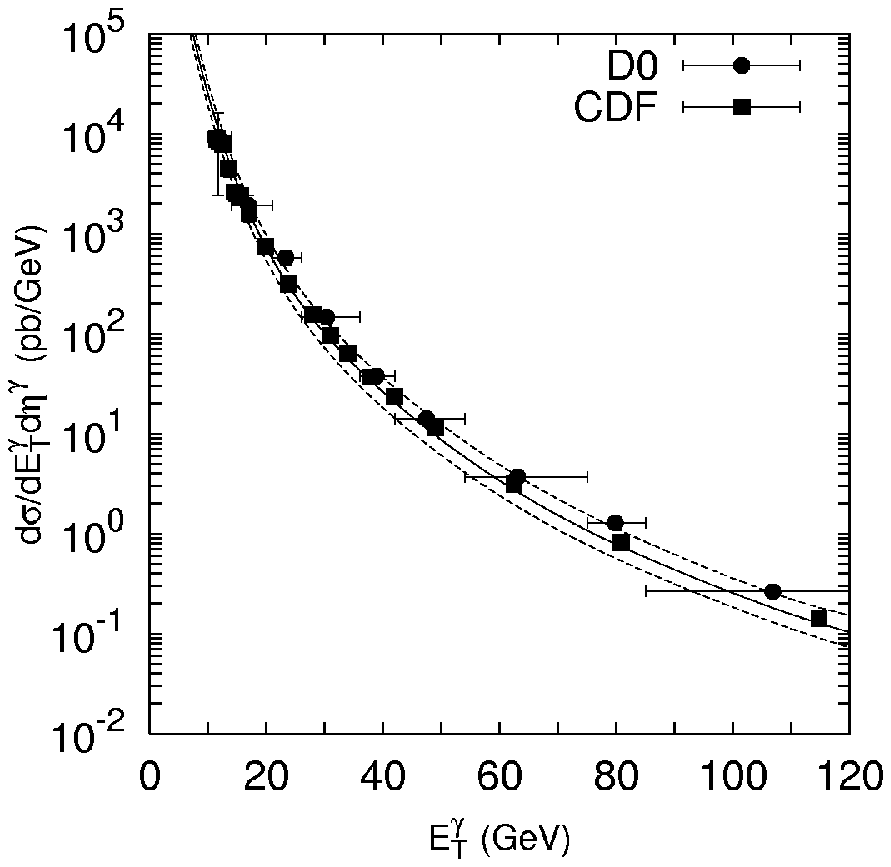, width = 22cm}
\caption{The double differential cross section $d\sigma/d E_T^\gamma d\eta^\gamma$
for inclusive prompt photon hadroproduction 
at $|\eta^\gamma| < 0.9$ 
and $\sqrt s = 1800$ GeV. All curves are the same as in Figure~3. 
The experimental data are from D$\oslash$~[1] and CDF~[3].}
\label{fig5}
\end{figure}

\newpage

\begin{figure}
\epsfig{figure=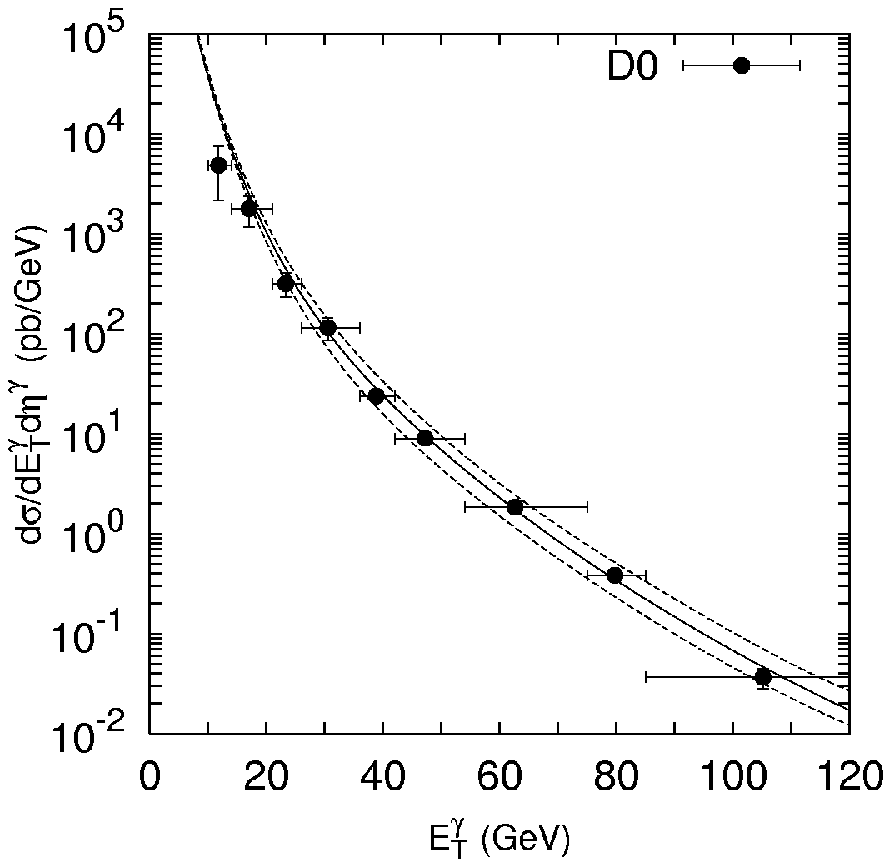, width = 22cm}
\caption{The double differential cross section $d\sigma/d E_T^\gamma d\eta^\gamma$
for inclusive prompt photon hadroproduction 
at $1.6 < |\eta^\gamma| < 2.5$ 
and $\sqrt s = 1800$ GeV. All curves are the same as in Figure~3. 
The experimental data are from D$\oslash$~[1].}
\label{fig6}
\end{figure}

\newpage

\begin{figure}
\epsfig{figure=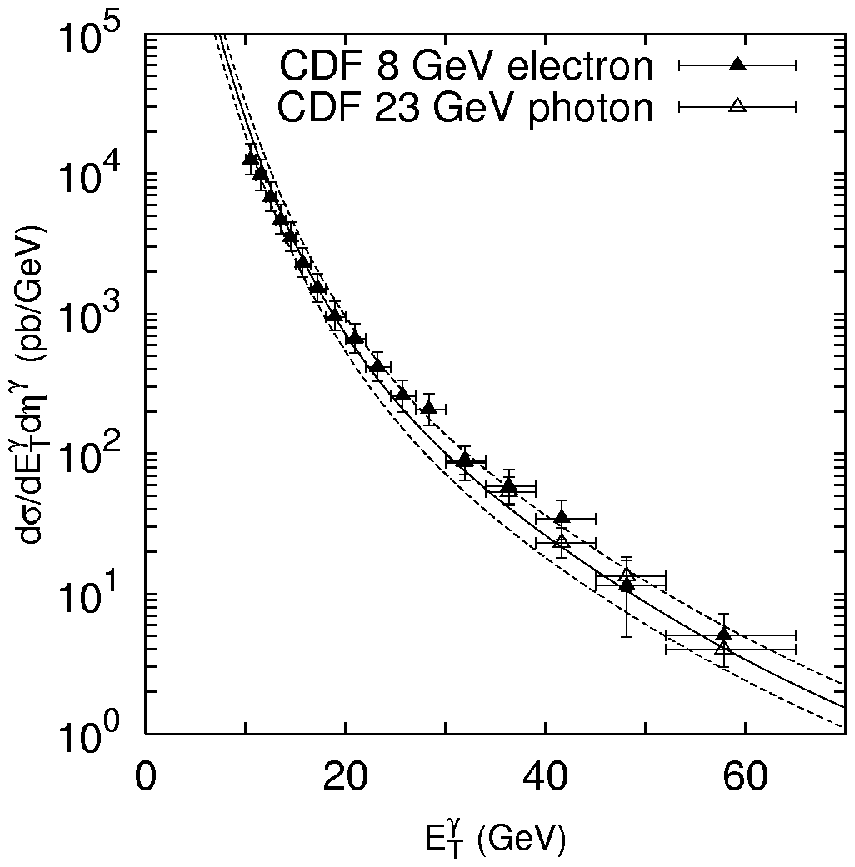, width = 22cm}
\caption{The double differential cross section $d\sigma/d E_T^\gamma d\eta^\gamma$
for inclusive prompt photon hadroproduction 
at $|\eta^\gamma| < 0.9$ 
and $\sqrt s = 1800$ GeV. All curves are the same as in Figure~3. 
The experimental data are from CDF~[4].}
\label{fig7}
\end{figure}

\newpage

\begin{figure}
\epsfig{figure=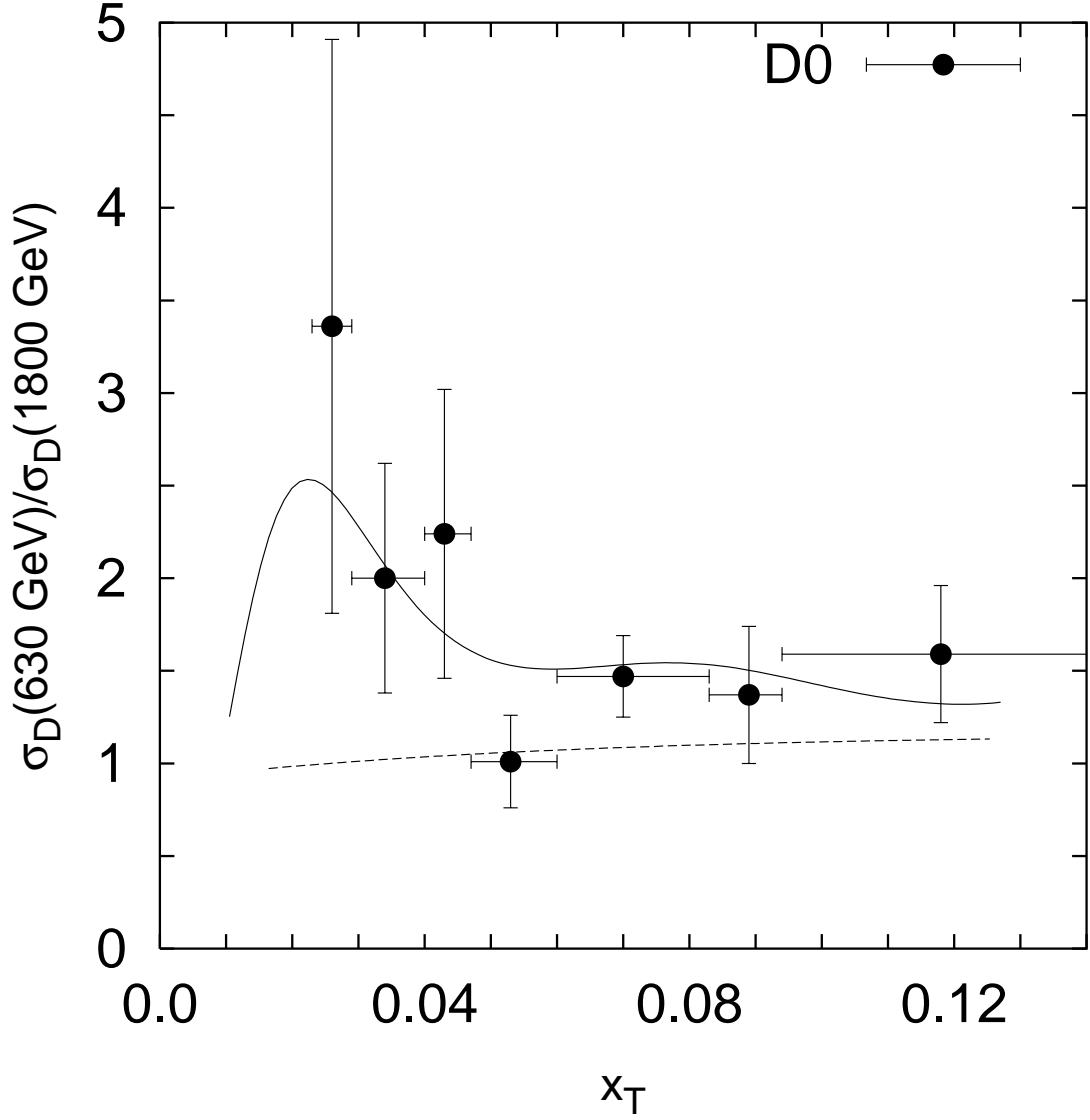, width = 22cm}
\caption{The ratio of the dimensionless cross sections
$\sigma_D(630\,{\rm GeV})/\sigma_D(1800\,{\rm GeV})$ as a
function of scaling variable $x_T$ at $|\eta^\gamma| < 0.9$. 
The solid line was obtained in the
$k_T$-factorization approach whereas dashed line corresponds to the 
collinear leading-order QCD calculations with the GRV parton 
densities of the proton. The renormalization and 
factorization scales are $\mu = E_T^\gamma$.
The experimental data are from D$\oslash$~[2].}
\label{fig8}
\end{figure}

\newpage

\begin{figure}
\epsfig{figure=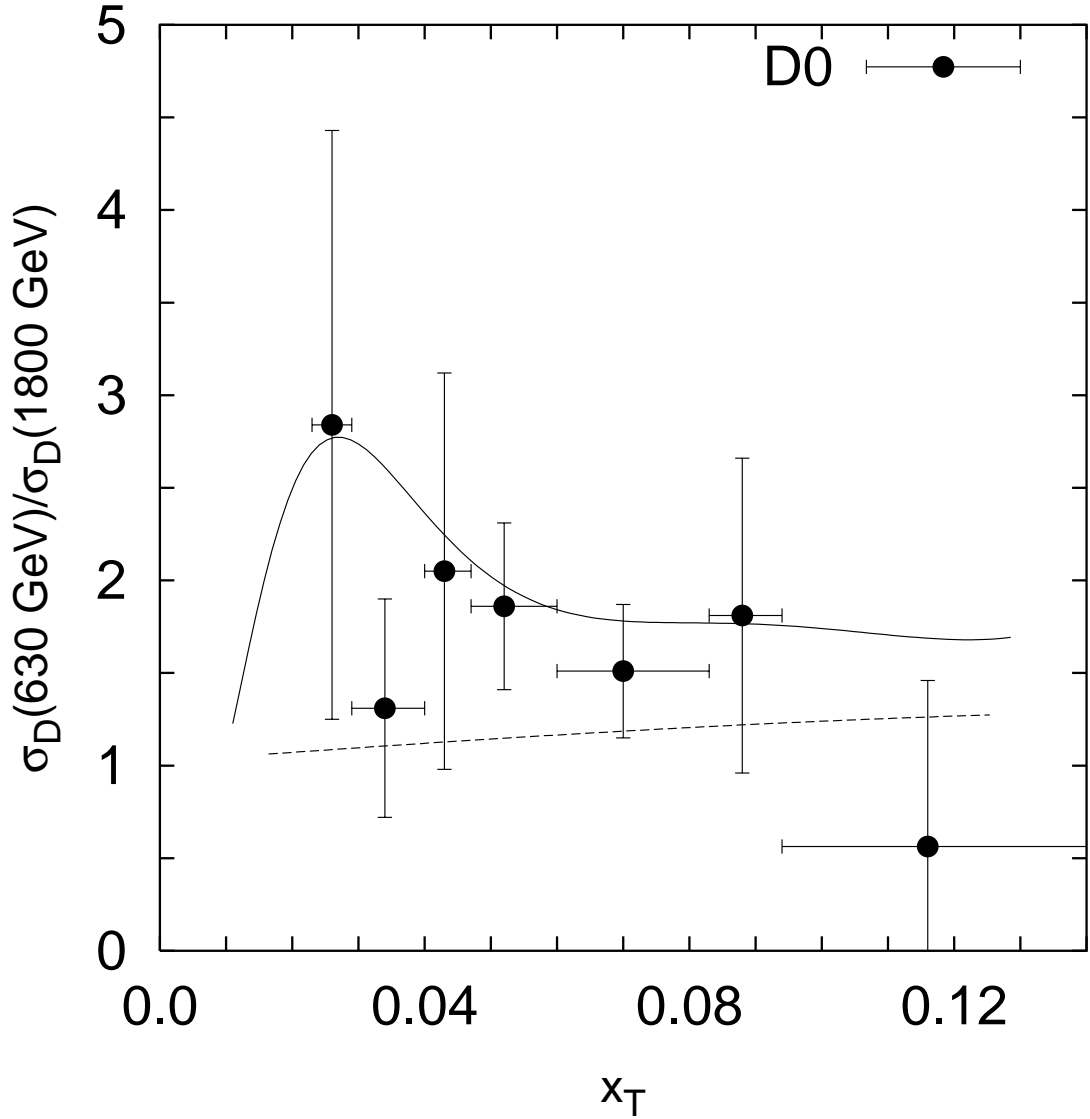, width = 22cm}
\caption{The ratio of the dimensionless cross sections
$\sigma_D(630\,{\rm GeV})/\sigma_D(1800\,{\rm GeV})$ as a
function of scaling variable $x_T$ at $1.6 < |\eta^\gamma| < 2.5$.
All curves are the same as in Figure~8. 
The experimental data are from D$\oslash$~[2].}
\label{fig9}
\end{figure}

\newpage

\begin{figure}
\epsfig{figure=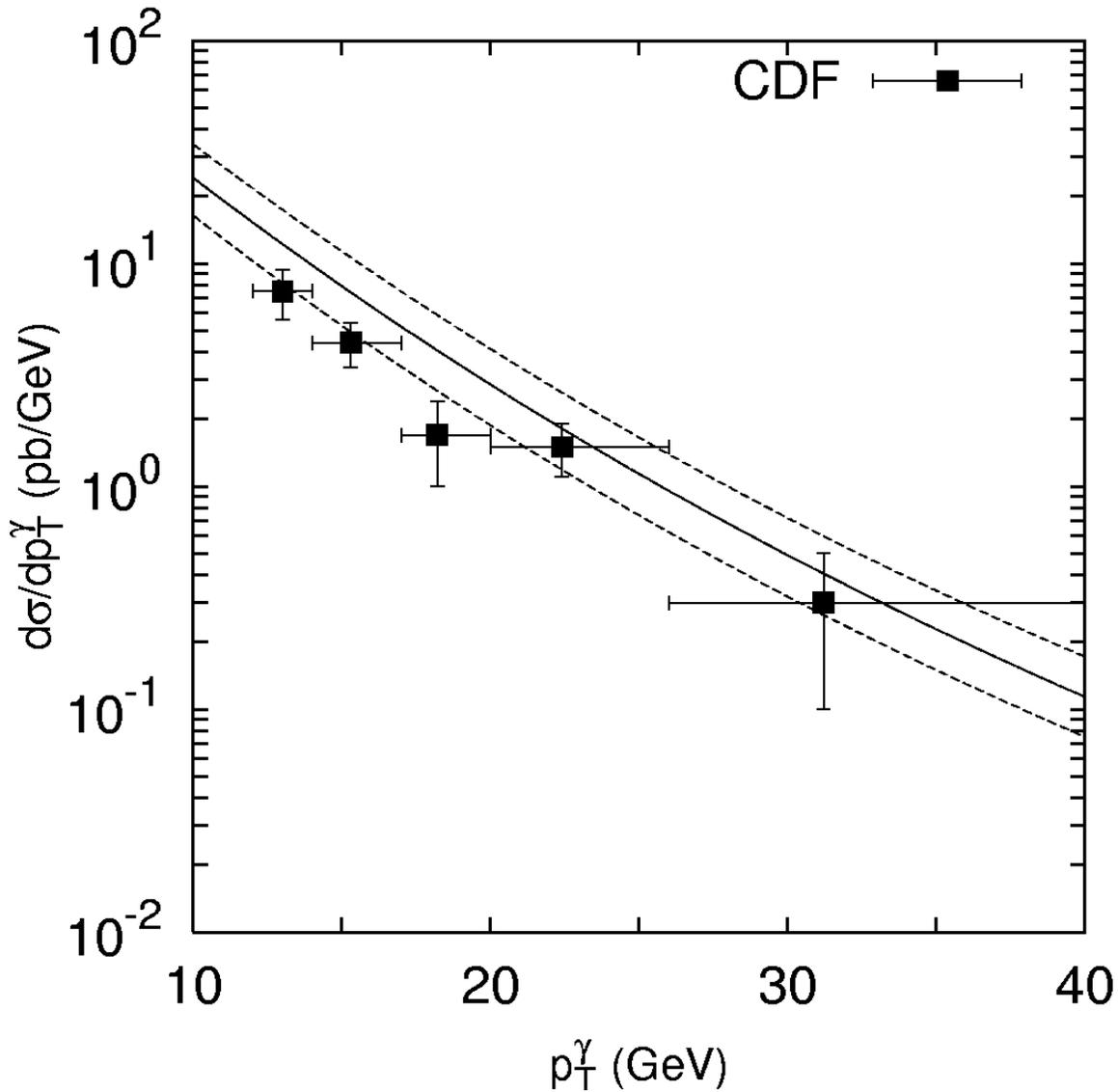, width = 22cm}
\caption{The differential cross section $d\sigma/d p_T^\gamma$
for associated prompt photon and muon hadroproduction at 
$|\eta^\gamma| < 0.9$, $|\eta^\mu| < 1.0$, $p_T^\mu > 4$ GeV and
$\sqrt s = 1800$ GeV. All curves are the same as in Figure~3. 
The experimental data are from CDF~[5].}
\label{fig10}
\end{figure}

\newpage

\begin{figure}
\epsfig{figure=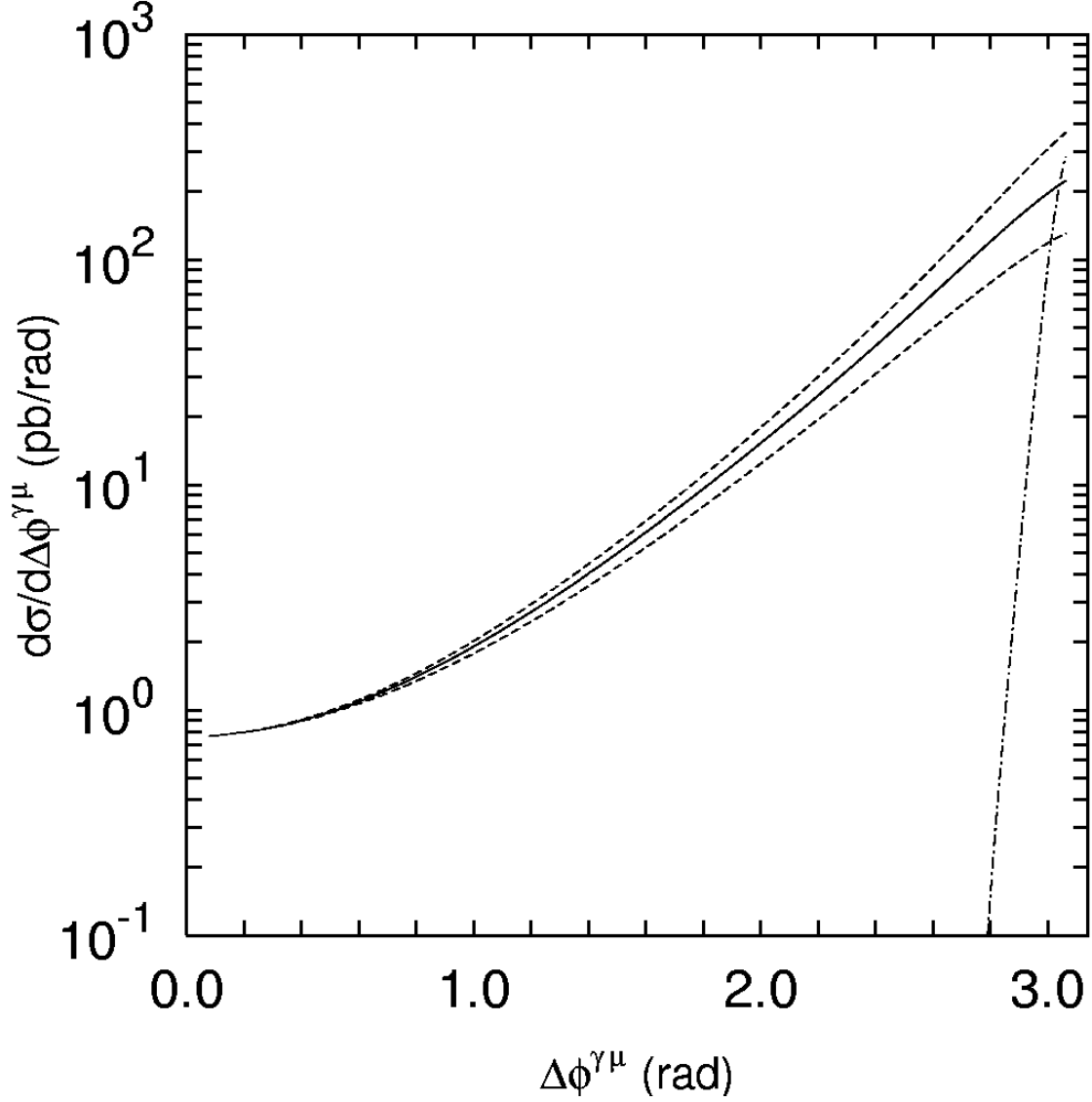, width = 22cm}
\caption{Azimuthal correlations in associated
prompt photon and muon hadroproduction at $|\eta^\gamma| < 0.9$, 
$|\eta^\mu| < 1.0$, $p_T^\mu > 4$ GeV and $\sqrt s = 1800$ GeV.
The solid line corresponds to the default scale $\mu = E_T^\gamma$, 
whereas upper and lower dashed lines correspond to the 
$\mu = E_T^\gamma/2$ and $\mu = 2 E_T^\gamma$ scales, respectively.
The dash-dotted line correspond to the sole LO QCD calculations
at $\mu = E_T^\gamma$.}
\label{fig11}
\end{figure}

\newpage

\begin{figure}
\epsfig{figure=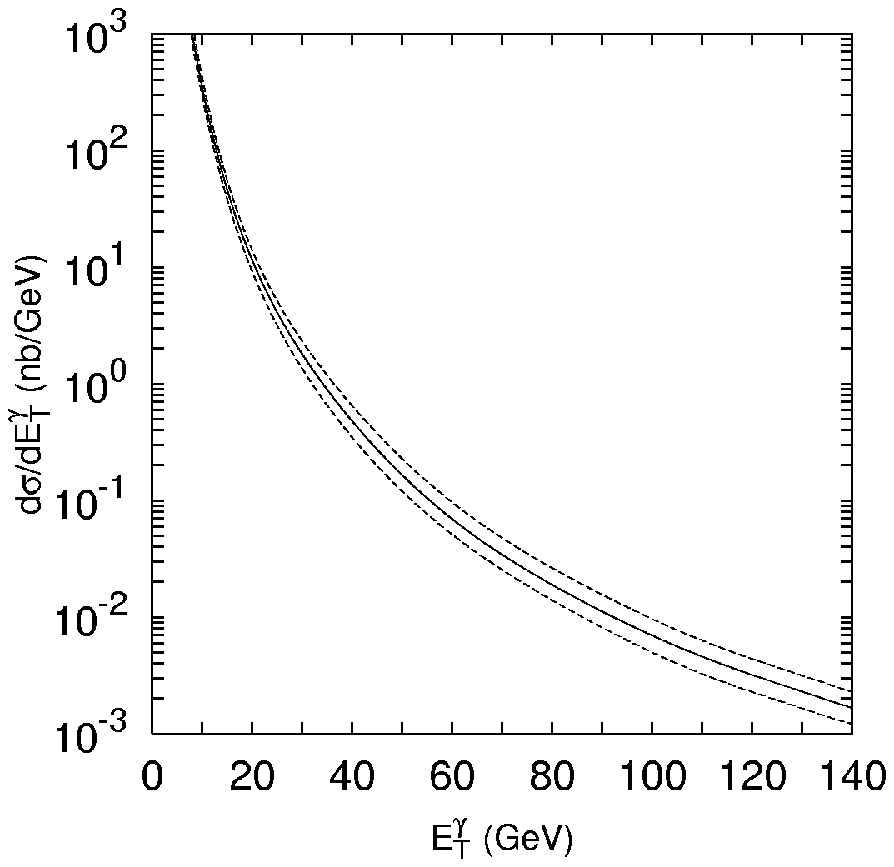, width = 22cm}
\caption{The differential cross section $d\sigma/d E_T^\gamma$
for inclusive prompt photon hadroproduction 
at $|\eta^\gamma| < 2.5$ and $\sqrt s = 14$ TeV. 
All curves are the same as in Figure~3.}
\label{fig12}
\end{figure}

\newpage

\begin{figure}
\epsfig{figure=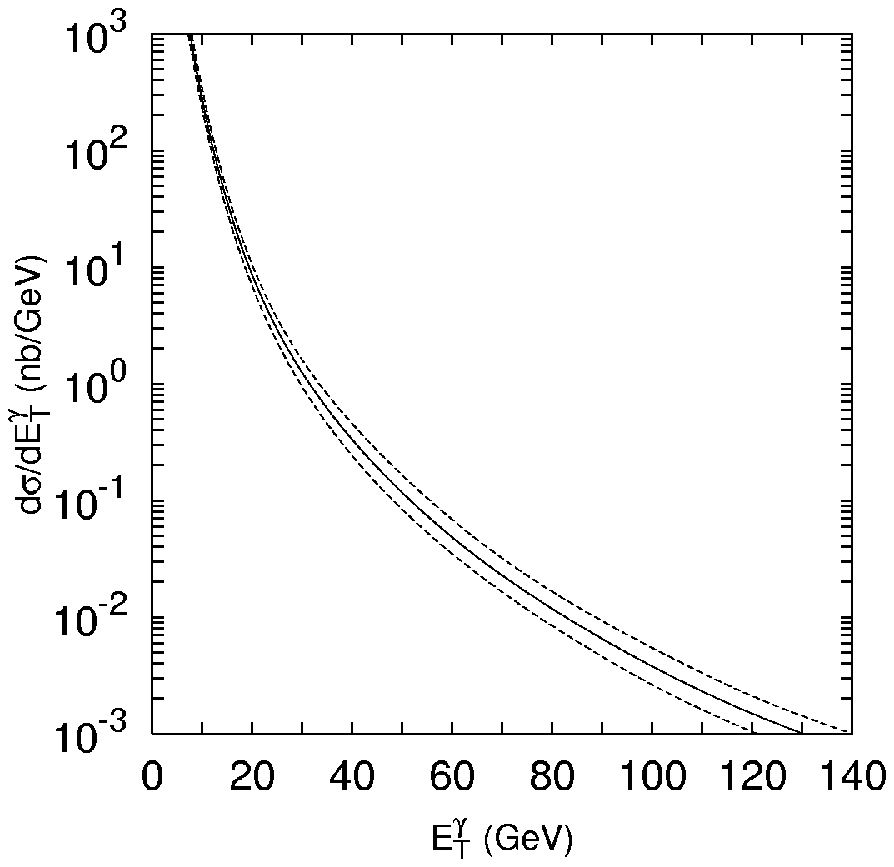, width = 22cm}
\caption{The differential cross section $d\sigma/d E_T^\gamma$
for inclusive prompt photon hadroproduction 
at $2.5 < |\eta^\gamma| < 4$ and $\sqrt s = 14$ TeV. 
All curves are the same as in Figure~3.}
\label{fig13}
\end{figure}

\end{document}